\documentclass[aip, jmp, 12pt]{revtex4-1}

\usepackage{amsmath}
\usepackage{amsthm}
\usepackage{amssymb}
\usepackage{graphicx}

\def\qe{\frac{q}{4 \pi \epsilon_0}}

\def\ad{\dot{\alpha}}

\def\atheta{\alpha_{\theta}}
\def\aphi{\alpha_{\phi}}

\def\f{F}

\def\real{\mathbb{R}}

\def\a{A}
\def\adual{\tilde{A}}

\def\v{V}

\def\vdual{\tilde{V}}

\def\m{\mathcal{M}}
\def\c{C}

\def\cddot{\ddot{C}}
\def\cdddot{\dddot{C}}
\def\cdot{\dot{C}}

\def\Vact(#1){{\left\langle #1 \right\rangle}}

\def\Pte(#1){_{#1}}
\def\Ptle(#1){|_{#1}}
\def\PtLe(#1){\Big|_{#1}}

\def\SigmaT{\Sigma_{\textup{T}}}
\def\Pdot{\dot{\mathrm{P}}}
\def\P{\mathrm{P}}

\def\ord{\mathcal{O}}
\def\const{\kappa}
\def\tDirac{{\textup{D}}}
\begin{document}
\title{The origin of the Schott term in the electromagnetic self force of a classical point charge}
\author{Michael R. Ferris and Jonathan Gratus }
\affiliation{Physics Department, Lancaster University, LA1 4YB, UK \\\& The Cockcroft Institute, UK}
\begin{abstract} The Schott term is the third order term in the
  electromagnetic self force of a charged point particle. The self force may be
  obtained by integrating the electromagnetic stress-energy-momentum
  tensor over the side of a narrow hypertube enclosing a section of worldline.
  This calculation has been repeated many times using two
  different hypertubes known as the Dirac Tube and the Bhabha Tube,
  however in previous calculations using a Bhabha Tube the Schott term does
  not arise as a result of this integration. In order to regain the
  Lorentz-Abraham-Dirac equation many authors have added an ad hoc
  compensatory term to the non-electromagnetic contribution to the
  total momentum. In this article the Schott term is obtained by direct integration of
  the electromagnetic stress-energy-momentum tensor.
\end{abstract}

\pacs{41.60.-m, 41.75.Ht, 41.90.+e}

\begin{flushright}
Cockcroft-11-15
\end{flushright}
\vspace{3em}

\maketitle

\section{Introduction}

The self force on an accelerating charged
particle is the force due to the particles own electromagnetic field. The
method used to derive the self force on a single particle depends on
the model on which the particle is based. In 1938 Dirac\cite{Dirac38} proposed a method
based on the point particle model. In this approach the
self force can be calculated directly from the
electromagnetic stress-energy-momentum tensor associated with the
Li\'{e}nard-Wiechert field, and is found to be the sum of three terms.
Two of these terms are finite, of which one is known as the
`radiation reaction term' and the other is known as the
`Schott term'. The third term is infinite. It is customary to treat this
infinite term as an electromagnetic contribution to mass, which combined
with the `bare mass' gives the observed mass of the particle.
The combining of electromagnetic and bare masses is known as
`mass renormalization' and leads to a finite expression for the
mass renormalized self force, which is the correction term in the equation of motion.

Let $C:I\subset\real\to\m$ be the proper time parameterized inextendible worldline of a point particle of
mass $m$, bare mass $m_0$ and charge $q$ where $\m$ is Minkowski
spacetime with metric $g$ of signature $(-,+,+,+)$ and Levi-Civita connection $\nabla$ where
%[
\begin{align}
\cdot=\c_{\ast}(d/d \tau),\qquad \cddot=\nabla_{\cdot}\cdot, \qquad \cdddot=\nabla_{\cdot} \nabla_{\cdot} \cdot
\end{align}
%]
and $\tau\in I$.  We use the SI unit convention but with the speed of light $c=1$. It follows
%[
\begin{align}
g(\cdot,\cdot)=-1,
\label{gCdCd}
\end{align}
%]
and hence
%[
\begin{align}
g(\cdot, \cddot)=0,
\qquad \textup{and} \qquad
g(\cdot, \cdddot)=-g(\cddot, \cddot).
\label{g_Cd_Cdd}
\end{align}
%]

\begin{figure}
\setlength{\unitlength}{1cm}
\centerline{\fbox{
\begin{picture}(14, 10)
\put(0.25, 0){\includegraphics[height=10cm, width=13.25cm]{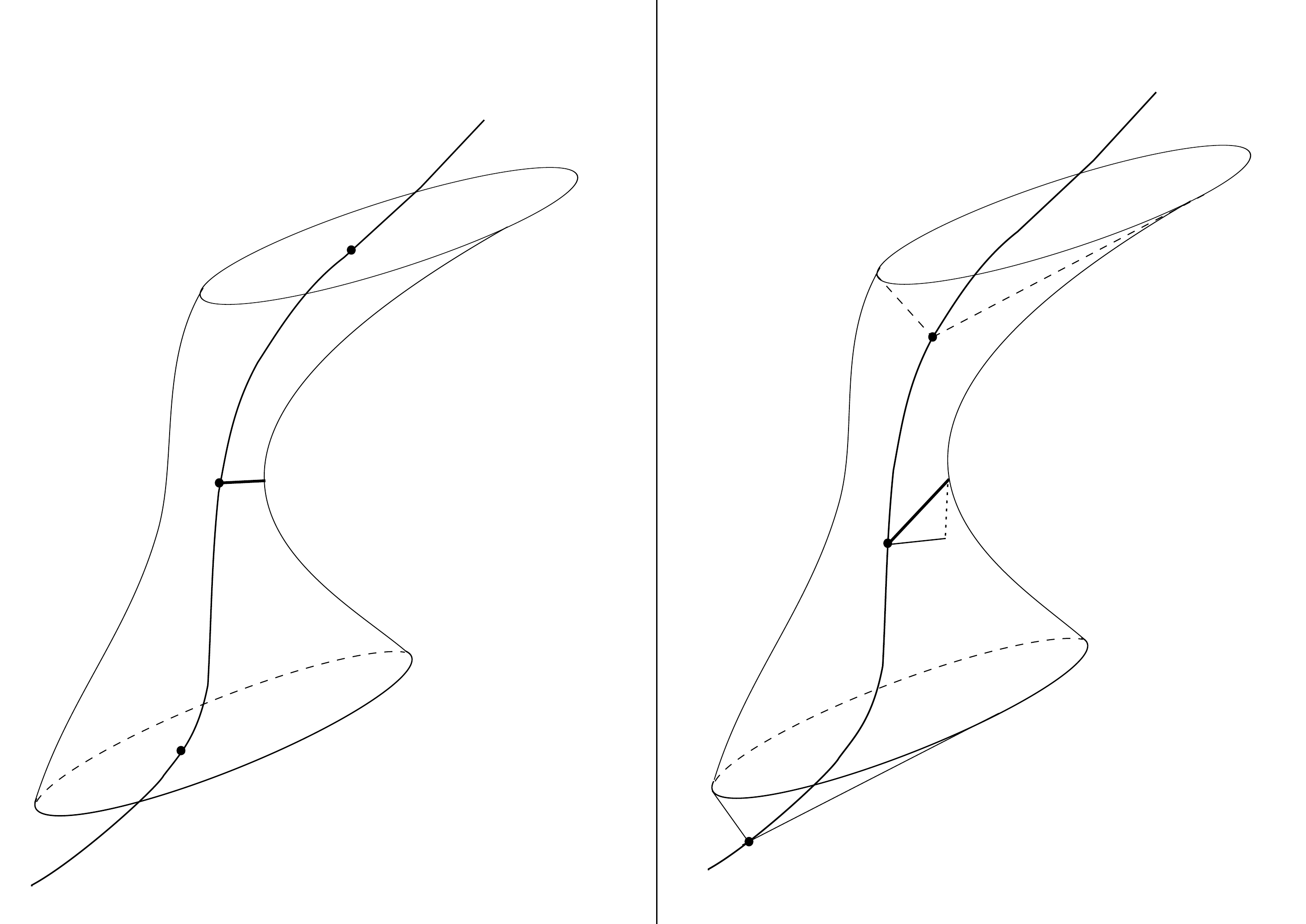}
}
\put(2.5, 9.5){Dirac Tube}
\put(9.5, 9.5){Bhabha Tube}
\put(4.4, 7.5){$\Sigma_2^\tDirac$}
\put(11.8, 7.7){$\Sigma_2$}
\put(2.5, 2.1){$\Sigma_1^\tDirac$}
\put(9.8, 2.2){$\Sigma_1$}
\put(0.8, 3.5){$\Sigma_{\textup{T}}^\tDirac$}
\put(8, 3.5){$\Sigma_{\textup{T}}$}
\put(2.3,4.3){$R_0^{\textup{D}}$}
\put(3,4.7){$Y$}
\put(9.7,3.8){$R_0$}
\put(10.3,4.7){$X$}
\put(1.9,4.7){$\tau_0$}
\put(9.1,4){$\tau_0$}
\put(1.5,1.85){$\tau_1$}
\put(8.1,.7){$\tau_1$}
\put(3.3,7.3){$\tau_2$}
\put(9.5,6.3){$\tau_2$}
\end{picture}
}}
\caption{The Dirac and Bhabha Tubes}
\label{fig_Tubes}
\end{figure}
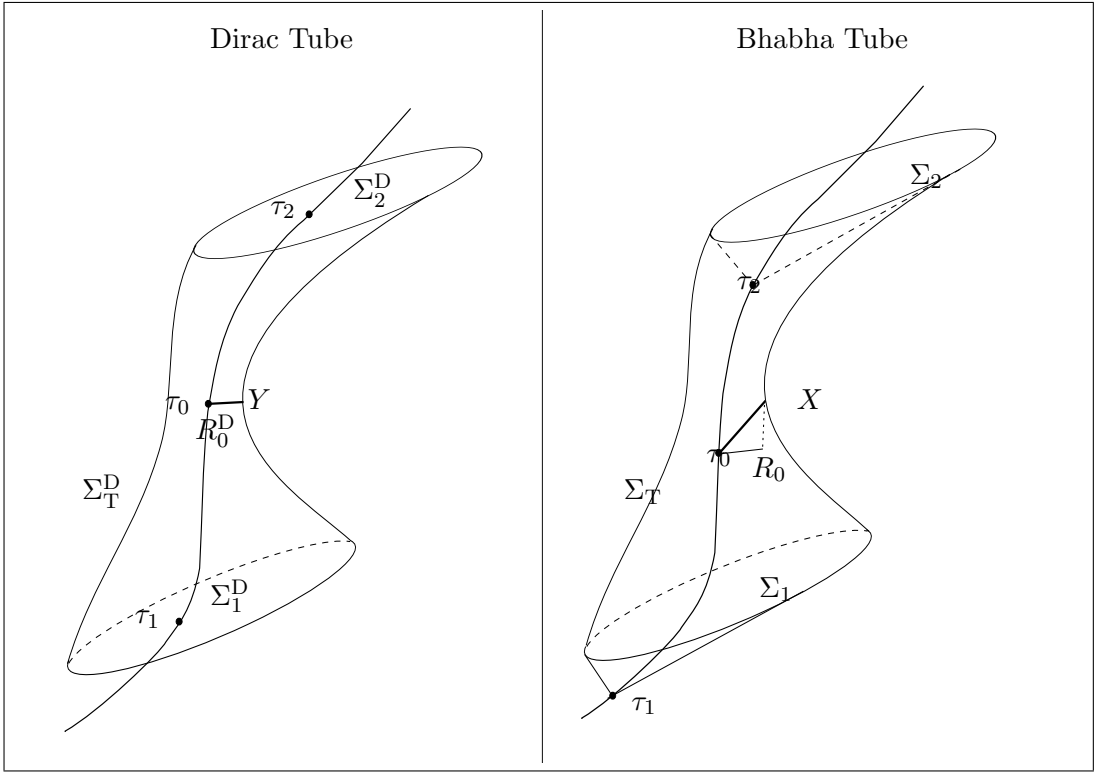

Within the point model framework the instantaneous electromagnetic
4-momentum arises as an integral of the electromagnetic
stress-energy-momentum tensor over a suitable $3$-surface in
spacetime. In Dirac's calculation the surface is the side
$\Sigma^\tDirac_T$ of a thin tube, of spatial radius
$R_0^\tDirac$, enclosing a section of the worldline $C$. See
FIG. \ref{fig_Tubes}. Since the displacement vector $Y$ defining
 the Dirac tube is spacelike, the Li\'{e}nard-Wiechert potential
is written as a series expansion in proper time.  When using
a Dirac tube the integration of the electromagnetic stress-energy-momentum tensor
gives for the self force \cite{Dirac38,Teitelboim70,Galtsov02,Norton}
%[
\begin{align}
f^{\textup{D}}_{\textup{self}}=
\kappa\:\Big(\:\tfrac23\big( \cdddot-g(\cddot, \cddot)\cdot\big)
-\lim_{R^\tDirac_0 \rightarrow 0}\frac{1 }{2 R^\tDirac_0}\cddot\Big),
\label{f_self}
\end{align}
%]
where
%[
\begin{align}
\kappa=\frac{q^2}{4 \pi \epsilon_0}.
\label{def_kappa}
\end{align}
%]

The first term is known as the `Schott term' and the second term is
sometimes called the `radiation reaction' term. The third term is the
singular term whose coefficient will later be identified as an
electromagnetic mass.

An alternative approach, proposed by Bhabha\cite{Bhabha39} in 1939 ,
is to integrate the electromagnetic stress-energy-momentum tensor over the side
$\Sigma_{\textup{T}}$ of the Bhabha tube with spatial radius $R_0$. The
principal advantage of this approach is that the displacement vector
$X$ is lightlike and as a result the Li\'{e}nard-Wiechert potential, and the corresponding electromagnetic field and stress-energy-momentum
tensor, can written explicitly. However previous articles which use a
Bhabha tube to evaluate the self force give the following expression
\cite{Norton, Bhabha39,Poisson99,Tucker06,Parrott86}
%[
\begin{align}
f_{\textup{self}}^{\textup{B}}&=-\kappa\:\Big(\:\tfrac23g(\cddot,
\cddot)\cdot+\lim_{R_0 \rightarrow 0}\frac{1 }{2 R_0}\cddot\Big)
=
f^{\textup{D}}_{\textup{self}}-\tfrac23\kappa\cdddot.
\label{f_self_wrong}
\end{align}
%]
Thus the Schott term is missing, indicating a major drawback of these
approaches.

In 2006 Gal'tsov and Spirin \cite{Galtsov02} draw attention to this discrepancy. They claim the Schott term should
arise directly from the electromagnetic stress-energy-momentum tensor and provide a derivation using Dirac's space-like coordinate system in order to show this. However they propose the missing term in \eqref{f_self_wrong} is a consequence of the light-like coordinates used to define the Bhabha tube. We show the term may be obtained using light-like coordinates and therefore that the missing term results from the procedure followed and not from the nature of the coordinates.

Calculation of the self force requires a minimum of two limits to be taken; the
shrinking of the hypertube onto the worldline and the bringing together of the caps.
In previous calculations \cite{Norton, Bhabha39,Poisson99,Tucker06,Parrott86} the Schott term remains
unnoticed because the former limit is taken before the latter.
In this article the calculation of the self force requires
three limits to be taken, the shrinking of the Bhabha tube $\SigmaT$ onto the
worldline $\c$ i.e. $R_0\to 0$,  and the bringing together of the lightlike caps $\Sigma_1$ and $\Sigma_2$
onto the lightlike cone with vertex $\c(\tau_0)$ i.e. $\tau_1\to\tau_0$  $\tau_2\to\tau_0$, where $\tau_0$ is the proper time of the point
where we wish to evaluate the self force (see FIG.\ref{fig_Tubes}).
We therefore have the freedom to choose the order of these limits.  We choose to let the three limits take place simultaneously,
subject to the constraint that
%[
\begin{align}
\lambda=\raisebox{0.4cm}{$\displaystyle{\lim_{\substack{R_0
        \rightarrow 0\\\tau_1 \rightarrow \tau_0\\\tau_2 \rightarrow
        \tau_0}}}$}
\Big(  \frac{\tau_1+\tau_2-2\tau_0}{4R_0}\Big)
\label{def_lambda}
\end{align}
%]
where $\lambda\in\real$ is finite. This gives the self force as
%[
\begin{align}
f_{\textup{self}}
&=
-\const\Big(
\tfrac{2}{3}g(\cddot, \cddot)\cdot
+\lambda\cdddot
+ \lim_{R_0 \rightarrow 0} \frac{1}{2R_0}\cddot
\Big)
\label{Intr_Pdot_res_expan}
\end{align}
%]
which is in agreement with $f^{\textup{D}}_{\textup{self}}$ if
$\lambda=-\tfrac23$, hence the Schott term arises by direct integration of the electromagnetic stress-energy-momentum tensor.

We suppose a balance of momentum
%[
\begin{align}
\dot{P}_{\text{PART}}  +\dot{P}_{\text{EM}}=f_{\text{ext}}
\label{Ppart_PEM}
\end{align}
%]
where total momentum has been separated into electromagnetic contribution $P_{\text{EM}}$ and non-electromagnetic
contribution $P_{\text{PART}}$, and $\dot{P}=\nabla_{\cdot}P$. All the external forces acting on the particle are denoted by $f_{\text{ext}}$, and in the following we show  $\dot{P}_{\text{EM}}=-f_{\textup{self}}$.
A suitable choice for the non-electromagnetic momentum
$P_{\text{PART}}$ has to be made.
Most external forces $f_{\text{ext}}$, including the Lorentz force,
 are orthogonal to $\cdot$:
%[
\begin{align}
g(f_{\text{ext}},\cdot)=0.
\label{f_ext_orthog}
\end{align}
%]
For such an external force, if (\ref{Intr_Pdot_res_expan}) is obtained then a natural choice for $P_{\text{PART}}$ is
%[
\begin{align}
P_{\text{PART}}=m_0  \cdot.
\label{P_not_inc_Schott}
\end{align}
%]
Combining (\ref{Intr_Pdot_res_expan}), (\ref{Ppart_PEM}) and
(\ref{P_not_inc_Schott}) gives
%[
\begin{equation}
\begin{aligned}
m_0  \cddot &= f_{\text{ext}}+f_{\textup{self}}\\
 &= f_{\text{ext}}-\const\Big(
\tfrac{2}{3}g(\cddot, \cddot)\cdot
+\lambda\cdddot\Big)
- \lim_{R_0 \rightarrow 0} \frac{\const}{2R_0}\cddot.
\end{aligned}
\label{balance}
\end{equation}
%]
Thus we satisfy the orthogonality condition (\ref{g_Cd_Cdd}) provided
$\lambda=-\tfrac23$. By contrast, if (\ref{f_self_wrong}) is obtained one cannot set
%[
\begin{align*}
m_0\cddot = f_{\text{ext}} + f_{\textup{self}}^{\text{B}}
\end{align*}
%]
and satisfy (\ref{g_Cd_Cdd}). Instead an extra term is added ad hoc
to the non-electromagnetic contribution to the force in order to
compensate for the missing Schott term \cite{Norton,Bhabha39,Poisson99,Tucker06}:
%[
%[
\begin{align}
\dot{P}^{\textup{B}}_{\text{PART}}=m_0  \cddot + \tfrac23\kappa \cdddot.
\label{P_inc_Schott}
\end{align}
%]
%]

With $\lambda=-\tfrac23$ equation (\ref{balance}) gives
%[
\begin{align*}
m_0  \cddot + \lim_{R_0 \rightarrow 0} \frac{\const}{2R_0}\cddot = f_{\text{ext}}+\tfrac23 \const\Big(\cdddot
-g(\cddot, \cddot)\cdot
\Big).
\end{align*}
%]
The singular coefficient is identified as an electromagnetic contribution
 to the total mass of the particle,  such that the observed mass m satisfies
%[
\begin{align}
 m =m_0 +\lim_{R_0 \rightarrow 0}\frac{\kappa }{2 R_0}.
\label{def_mass_bare}
 \end{align}
%]
The resulting equation of motion for a point charge in an external field
is the Lorentz-Abraham-Dirac equation
%[
\begin{align}
m \cddot=
{f_{\text{ext}}}+\frac{q^2}{6\pi \epsilon_0 }(\cdddot-g(\cddot, \cddot)\cdot  ).
\label{AbramLor_Dir_eqn}
\end{align}
%]

%%%%%%%%%%%%%%%%%%%%%%%%%%%%%%%%%%%%%%%%%%%%%%%%%%%%%%%%%%%%%%%%%%%%%%
\section{Calculation of the self force}
In the following calculation expression~\eqref{Intr_Pdot_res_expan} for the self force is obtained by direct integration of the the electromagnetic
stress-energy-momentum tensor over the side $\SigmaT$ of the Bhabha Tube, and hence the ad hoc term in \eqref{P_inc_Schott} is avoided in the derivation of the Lorentz-Abraham-Dirac equation.

We use a global Lorentzian frame $(x^0, x^1, x^2, x^3)$ on Minkowski
spacetime $\m$ with metric
%[
\begin{align*}
g= - dx^0\otimes dx^0+dx^1 \otimes dx^1 +dx^2 \otimes dx^2 +dx^3 \otimes dx^3.
\end{align*}
%]
In the following implicit summation is over $j, k, l=0,\ldots,3$.

All fields will be regarded as sections of tensor bundles over
appropriate domains of $\m$. Sections of the tangent bundle over $\m$
will be denoted $\Gamma T \m$ while sections of the bundle of exterior
$p$-forms will be denoted $\Gamma \Lambda^p \m$. Sections over spacetime
excluding the worldline are written $\Gamma T (\m\backslash C)$ and
$\Gamma\Lambda^p (\m\backslash C)$. For any vector field
$\v$ denote by $\widetilde{\v}$ the associated 1-form defined by
$\widetilde{\v} = g(\v,-)$. The operator $d$ will denote the exterior
derivative and $\displaystyle{i_\v}$ the contraction operator with
respect to $\v$. The operator $\star$ is the Hodge map associated with
metric $g$.

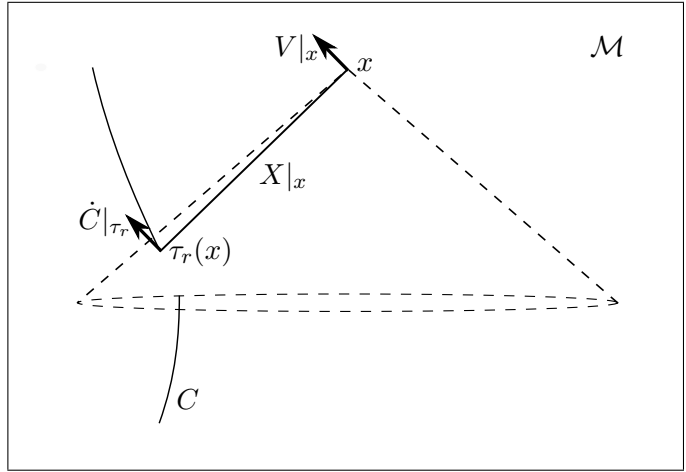
\begin{figure}[h]
\setlength{\unitlength}{1cm}
\centerline{\fbox{
\begin{picture}(8.5, 6)
\put(0, 0){\includegraphics[height=6cm, width=8.4cm]{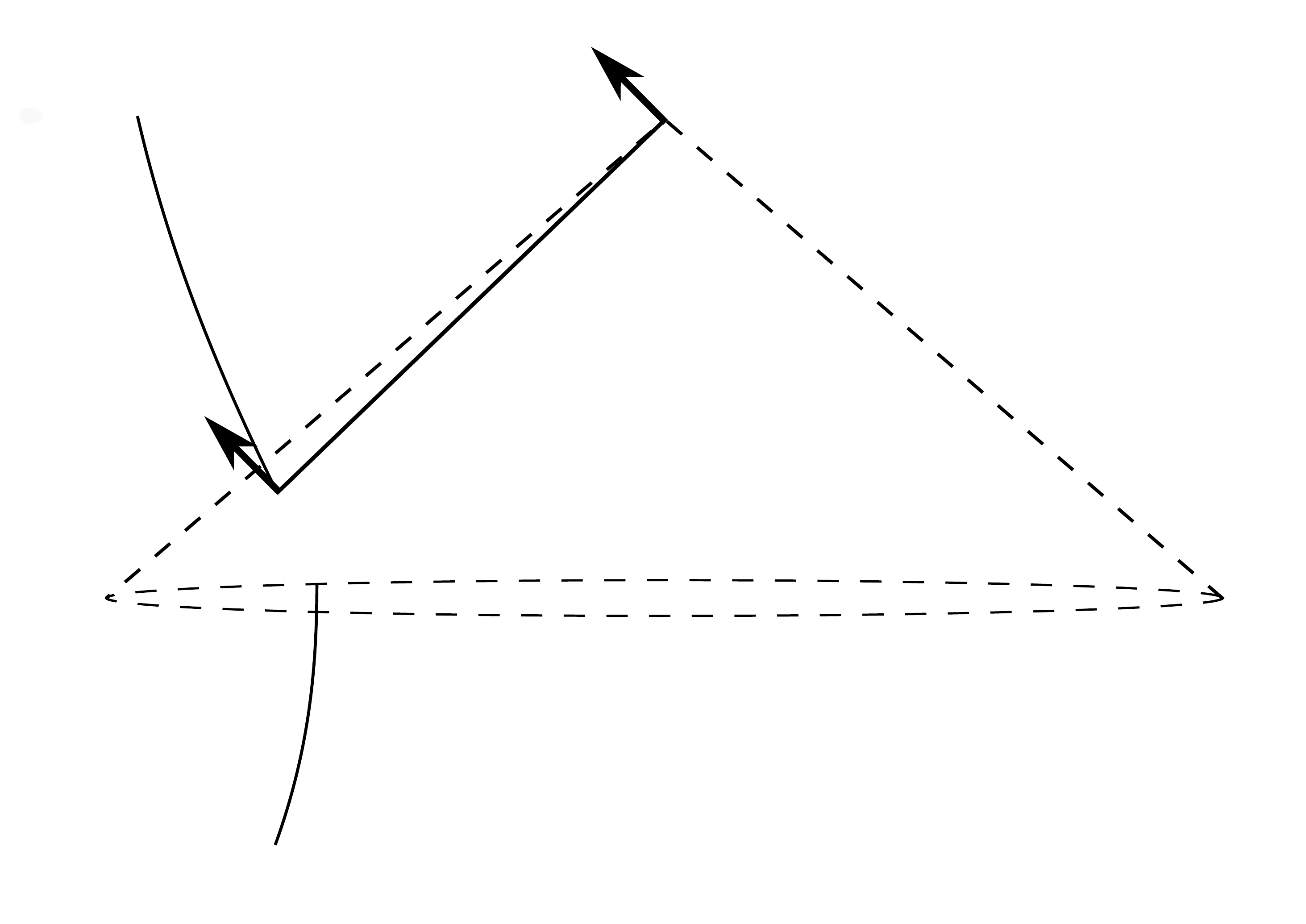}
}
\put(7.5, 5.4){$\m$}
\put(4.4, 5.2){$x$}
\put(3.1, 3.7){$X|_x$}
\put(2, 0.7){$\c$}
\put(1.9, 2.7){$\tau_r(x)$}
\put(0.7, 3.1){$\cdot|_{\tau_r}$}
\put(3.3, 5.4){$\v|_{x}$}
\end{picture}
}}
\caption{Retarded time $\tau_r(x)$ and spacetime vectors $\v|_x$ and $X|_x$.}
\label{fig_cone}
\end{figure}

The Bhabha tube, defined by the parameters $\tau_1,\tau_2\in I$ and $R_0>0$,
is given by $\displaystyle{\Sigma_1 \cup \Sigma_2 \cup \SigmaT}$ where
%[
\begin{equation}
\begin{aligned}
\Sigma_\mu &= \Big\{\c(\tau_\mu)+X\quad\Big|\quad g(X, X)=0,\quad -g(X,
\cdot)<R_0\Big\}\,,
\\
\SigmaT &= \Big\{\c(\tau)+X\quad\Big|\quad g(X,
X)=0,\quad-g(X, \cdot)=R_0, \quad\tau_1\leq\tau\leq\tau_2\Big\}
\end{aligned}
\label{Bhabha_Tube}
\end{equation}
%]
for $\mu=1,2$. Here we have used the affine structure of $\m$ to add a vector to a
point to give another point in $\m$.
For comparison the Dirac tube is given by
$\displaystyle{\Sigma^\tDirac_1 \cup
  \Sigma^\tDirac_2 \cup
  \Sigma^\tDirac_{\textup{T}}}$ where
%[
\begin{align*}
\Sigma^\tDirac_\mu &= \Big\{\c(\tau_\mu)+Y  \quad\Big|\quad g(Y,
\cdot)=0,\quad g(Y, Y)<(R^\tDirac_0)^2\Big\}\,,
\\
\Sigma^\tDirac_{\textup{T}} &= \Big\{\c(\tau)+Y \quad\Big|\quad g(Y, \cdot)=0\quad g(Y, Y)=(R^\tDirac_0)^2, \quad\tau_1\leq\tau\leq\tau_2\Big\}.
\end{align*}
%]
For every field point $x$ there is a unique point $\tau_r(x)$ at which the worldline intersects the retarded light-cone at $x$ (see FIG. \ref{fig_cone}),
%[
\begin{align}
\tau_r &: \mathcal{M}\backslash \c \rightarrow \real,\quad x \mapsto \tau_r(x).
\label{def_tau_R}
\end{align}
%]
The null vector $X \in \Gamma T (\mathcal{M}\backslash \c)$ is given by the difference between the field point \\
$x=(x^0, x^1, x^2, x^3)$ and the worldline point $\c(\tau_r(x))$
%[
\begin{align}
&X|_x= x-\c \big(\tau_r(x)\big), \quad g(X, X)=0.
\label{def_X}
\end{align}
%]
The vector fields $\v, \a, \dot{\a} \in \Gamma T (\mathcal{M}\backslash \c) $ are defined as
\begin{align}
\v|_x&=\cdot^j(\tau_r(x)) \frac{\partial}{\partial x^j},\quad\a|_x=\cddot^j(\tau_r(x))\frac{\partial}{\partial x^j}\quad \text{and} \quad\dot{\a}|_x=\cdddot^j(\tau_r(x)) \frac{\partial}{\partial x^j},
\label{def_V_A}
\end{align}
%]
hence from (\ref{g_Cd_Cdd})
%[
\begin{align}
g(\v, \v)=-1,\quad
g(\a, \v)=0,\quad \text{and}\quad
g(\dot{\a}, \v)=-g(\a, \a).
\label{orthog_V_A}
\end{align}
%]

The retarded Li\'{e}nard-Wiechert one-form potential\cite{Jackson99} $\Psi \in \Gamma
\Lambda^1 (\mathcal{M}\backslash\c)$ is given by
%[
\begin{align}
\Psi=\qe \frac{\vdual}{g(X, \v)}.
\label{def_Psi}
\end{align}
%]
The corresponding Li\'{e}nard-Wiechert electromagnetic field $\f \in \Gamma \Lambda^2 (\mathcal{M}\backslash\c)$ is given by
%[
\begin{align}
\f=d\Psi=\qe \frac{g(X, \v)\tilde{X}\wedge \adual-g(X, \a)\tilde{X} \wedge \vdual - \tilde{X}\wedge \vdual}{g(X, \v)^3}.
\label{def_F}
\end{align}
%]
It is sometimes beneficial in calculating the self force of a point charge to split $\f$ into a radiative and a bound contribution \cite{Teitelboim70, Poisson99, Galtsov02, Teitelboim80, Dirac38}. There are two common splittings which can be made that are independently motivated, however as emphasized by Gal'tsov and Spirin  \cite{Galtsov02} they give different radiative and bound contributions to the self force, differing by the Schott term. In this article we avoid splitting $\f$ because it is unnecessary in the calculation. 

The four electromagnetic stress $3-$forms\cite{BennTucker} $S_k\in \Gamma \Lambda^3
(\m\backslash\c)$ are given by
%[
\begin{align}
S_k=&\frac{\epsilon_0}{2}(i_{\partial/\partial x^k}
\f \wedge \star\f -i_{\partial/\partial x^k} \star\f  \wedge
\f ), \quad\quad\quad
\widetilde{\star S_k}=\textup{T}\Big(-,
\frac{\partial}{\partial x^k}\Big)
\label{def_Sk}
\end{align}
%]
where $\textup{T}$ is the symmetric stress-energy-momentum tensor
defined on $\m\backslash\c$.

It follows from the vacuum Maxwell equations that $S_k$ is closed in
$\m\backslash\c$, i.e. $d S_k=0$. Therefore in an arbitrary region $N
\subset \m \backslash \c$ of spacetime off the worldline
%[
\begin{align}
\int_{\partial \mathcal{N}} S_k = \int_{\mathcal{N}}d S_k=0.
\label{Stokes}
\end{align}
%]
The component of 4-momentum flux $\mathrm{P}_k^{(\Sigma)}\in\real$ through an arbitrary three dimensional
hypersurface $\Sigma\subset \m$ is defined by
%[
\begin{align}
\mathrm{P}_k^{(\Sigma)} &=\int_{\Sigma} S_k.
\label{def_Pk}
\end{align}
%]
The lightlike hypersurfaces $\Sigma_1$, $\Sigma_2$, and the timelike
hypersurface $\SigmaT$ (FIG. \ref{fig_Tubes}) are defined by
(\ref{Bhabha_Tube}) and let $\Sigma_1$ be negatively oriented with respect to $\SigmaT$ and $\Sigma_2$.

The instantaneous 4-momentum derivative at proper time $\tau_0 \in I$ is defined by
%[
\begin{align}
\Pdot_k (\tau_0)=\raisebox{0.4cm}{$\displaystyle{\lim_{\substack{R_0 \rightarrow 0\\\tau_1 \rightarrow \tau_0\\\tau_2 \rightarrow \tau_0}}}$}\:\bigg(\frac{1}{\tau_2-\tau_1}\P_k^{(\SigmaT)}\bigg)
\label{def_Pdot}
\end{align}
%]
where $\P_k^{(\SigmaT)}$ is given by (\ref{def_Pk}). This definition
is justified heuristically as follows.
Inspired by (\ref{Stokes}) we wish to
write
%[
\begin{align}
\P_k^{(\SigmaT)}=\P_k^{(\Sigma_2)}-\P_k^{(\Sigma_1)}
\label{def_Pk_diff}
\end{align}
%]
Ignoring the fact that $\P_k^{(\Sigma_1)}$ and $\P_k^{(\Sigma_2)}$ are
both infinite, we assert
%[
\begin{equation}
\begin{aligned}
\Pdot_k (\tau_0) = \raisebox{0.4cm}{$\displaystyle{\lim_{\substack{R_0 \rightarrow 0\\\tau_1 \rightarrow \tau_0\\\tau_2 \rightarrow \tau_0}}}$}\:\bigg(\frac{1}{\tau_2-\tau_1}\Big(\P_k^{(\Sigma_2)}-\P_k^{(\Sigma_1)}\Big)\bigg)
\end{aligned}
\label{def_Pk_dot}
\end{equation}
%]
Inserting (\ref{def_Pk_diff}) into (\ref{def_Pk_dot}) yields (\ref{def_Pdot}).

We define the vector $\Pdot_{\text{EM}}(\tau_0) \in  T_{\c(\tau_0)}\m$ by
%[
\begin{align}
\Pdot_{\text{EM}}(\tau_0)=\Pdot_k(\tau_0) g^{k l}\frac{\partial}{\partial x^l}
\label{def_Pdot_dual}
\end{align}
%]
where $g^{k l}=g^{-1}(dx^k, dx^l)$ and $g^{-1}$ is the inverse metric on $\m$.
Since $\tau_0$ is arbitrary there is an induced vector field $\Pdot_{\text{EM}}$ on the curve $\c$.
%%\subsection*{Coordinates}

To simplify the problem we introduce the retarded
`Newman-Unti' \cite{Newman}
coordinates $(\tau, R, \theta, \phi)$, where $\tau \in I$, $R>0$, $0<
\theta< \pi$ and $0< \phi< 2\pi$.
The coordinate transformation between $(\tau, R, \theta, \phi)$ and $(x^0, x^1, x^2, x^3)$ is given by:
%[
\begin{equation}
\begin{aligned}
x^0&=\c^{0}(\tau)+\frac{R}{\alpha},\\
x^1&=\c^{1}(\tau)+\frac{R}{\alpha}\sin(\theta)\cos(\phi),\\
x^2&=\c^{2}(\tau)+\frac{R}{\alpha}\sin(\theta)\sin(\phi),\\
x^3&=\c^{3}(\tau)+\frac{R}{\alpha}\cos(\theta)
\end{aligned}
\label{coords}
\end{equation}
%]
where $\alpha \in\Gamma \Lambda^0 \mathcal{M}$ is given by
%[
\begin{align}
\alpha(\tau, \theta, \phi) = \frac{g(X, \dot{\c}(\tau))}{g(X, \partial_{x^0})}
=\cdot^0(\tau) -\cdot^1 (\tau)\sin(\theta) \cos(\phi) - \cdot^2(\tau) \sin(\theta) \sin(\phi) -\cdot^3(\tau) \cos(\theta)
\label{def_alpha}
\end{align}
%]
From (\ref{coords}) and (\ref{def_alpha}) it follows
%[
\begin{align}
R=-g(X, \dot{\c}(\tau))
\quad\quad \textup{and}\quad\quad
\tau=\tau_r(x(\tau, R, \theta, \phi)).
\label{tau_taur_R}
\end{align}
%]
The spherical coordinates $\theta$ and $\phi$ are deduced from the
global Lorentzian frame $(x^0, x^1, x^2, x^3)$. In this coordinate system the metric is given by
%[
\begin{align*}
g =\Big(2 R \frac{\ad}{\alpha}-1\Big) d\tau \otimes d\tau
- (d\tau \otimes d R+ d R \otimes d\tau)
+ \frac{R^2}{\alpha^2}(d\theta \otimes d\theta
+  \sin^2(\theta)  d\phi \otimes d\phi)
\end{align*}
%]
and the vectors $X$, $\v$ and $\a$ are written
%[
\begin{equation}
\begin{aligned}
X&=R\frac{\partial}{\partial R},
\end{aligned}
\label{def_XX}
\end{equation}
%]
%[
\begin{equation}
\begin{aligned}
\v&=\frac{\partial}{\partial \tau}+\frac{\ad}{\alpha}R \frac{\partial}{\partial R}
\end{aligned}
\label{def_V}
\end{equation}
%]
and
%[
\begin{equation}
\begin{aligned}
\a&=\frac{\ad}{\alpha} \frac{\partial}{\partial \tau}+ \frac{\dot{\alpha}}{\alpha}\Big( \frac{R\dot{\alpha}}{\alpha}-1\Big) \frac{\partial}{\partial R} +\frac{\ad \aphi -\alpha \dot{\aphi}}{R\sin^2(\theta)} \frac{\partial}{\partial \phi} +\frac{\ad \atheta -\alpha \dot{\atheta}}{R} \frac{\partial}{\partial \theta}
\end{aligned}
\label{def_A}
\end{equation}
%]
where
%[
\begin{align*}
\ad=\frac{\partial \alpha}{\partial \tau}, \quad\atheta=\frac{\partial
  \alpha}{\partial \theta}, \quad\aphi=\frac{\partial \alpha}{\partial
  \phi}
\end{align*}
%]
In the coordinate system $\Sigma_{\textup{T}}$ is given by
%[
\begin{align}
\SigmaT &= \Big\{(\tau, R, \theta, \phi)\Big|\tau_1\leq\tau\leq\tau_2,\quad R=R_0,\quad 0\leq\theta\leq\pi,\quad 0\leq \phi\leq 2\pi\Big\}
\label{SigmaT_coords}
\end{align}
%]
%%\subsection*{Expansion around the momentarily comoving Lorentz frame }
Setting $\tau=\tau_0+\delta$ we expand $S_k$ in powers of $\delta$. We adapt the global Lorentz frame such that
%[
\begin{equation}
\begin{aligned}
&x^j(\c(\tau_0))=0\quad\text{for} \quad j=0, 1, 2, 3
\\
\text{and} \quad \quad&\cdot(\tau_0)= \frac{\partial}{\partial x^0},\quad\cddot(\tau_0)= a\frac{\partial}{\partial x^3},\quad \cdddot(\tau_0)=b^j \frac{\partial}{\partial x^j}
\end{aligned}
\label{def_x_i}
\end{equation}
%]
where $a,b^j \in\real$ are constants given by
%[
\begin{align}
a=\sqrt{g(\cddot(\tau_0), \cddot(\tau_0))}, \quad\quad
b^j=dx^j(\cdddot(\tau_0))
\label{def_ab}
\end{align}
%]
and from (\ref{g_Cd_Cdd})
%[
\begin{align*}
b^{0}=a^2
\end{align*}
%]
Thus expanding $\cdot$ and $\cddot$ we have
%[
\begin{equation}
\begin{aligned}
\cdot(\delta+\tau_0)
&=
\Big(1+ \frac{b^0}{2}\delta^2\Big)\frac{\partial}{\partial
  t}+\frac{b^1}{2}\delta^2\frac{\partial}{\partial x}+
\frac{b^2}{2}\delta^2\frac{\partial}{\partial y}+\Big(a \delta+
\frac{b^3}{2}\delta^2\Big)
\frac{\partial}{\partial z}+\ord(\delta^3),
\\
\cddot(\delta+\tau_0)
&=
b^0 \delta \frac{\partial}{\partial t}+b^1 \delta \frac{\partial}{\partial x}+b^2 \delta \frac{\partial}{\partial y}+\Big(a+b^3 \delta\Big) \frac{\partial}{\partial z}+\ord(\delta^2)
\end{aligned}
\label{cd_expansion}
\end{equation}
%]
From (\ref{def_V_A}) and (\ref{tau_taur_R}) we have
%[
\begin{align}
\v|_{(\delta+\tau_0,R,\theta,\phi)}
&=
\cdot(\delta+\tau_0)
\qquad\text{and}\qquad
\a|_{(\delta+\tau_0,R,\theta,\phi)}=\cddot(\delta+\tau_0)
\label{va_expand}
\end{align}
%]
It is useful to express $\v$ and $\a$ in mixed coordinates, with the
basis vectors in terms of the global Lorentz coordinates, but the
coefficients expressed in terms of the Newman-Unti coordinates.

%%\subsection*{Integration and limits}
We use the MAPLE script \cite{Maplescript, Manifolds} that accompanies this article
to evaluate the integral of $S_k$ over the side $\SigmaT$ of the Bhabha tube.
We substitute (28) and (29) into (16) in order to obtain an explicit
expression for the Li\'{e}nard-Wiechert potential $\Psi$ in Newman-Unti
coordinates (\ref{coords}). Taking the exterior derivative we obtain
the field $2-$form $\f$ and its Hodge dual $\displaystyle{\star\f}$.
We obtain expressions for the four translational Killing vectors
$\frac{\partial}{\partial x^k}$ in Newman-Unti coordinates and using
(\ref{def_Sk}) we obtain expressions for the four electromagnetic
stress $3-$forms $S_k$. Substituting the expansions (\ref{cd_expansion})
into these expressions we obtain the integrands, and finally using (\ref{SigmaT_coords})
 we integrate over $\SigmaT$. The result is

%[
\begin{equation}
\begin{aligned}
\frac{1}{\const}\int_{\SigmaT} S_{0}&= -\frac{1}{4}b^0 \frac{\delta_2^2-\delta_1^2}{R_0}-\frac{2}{3}a^2(\delta_2-\delta_1)-\frac{2}{3}a b^3(\delta_2^2-\delta_1^2)+\ord(\delta_1^3)+\ord(\delta_2^3),
\\
\frac{1}{\const}\int_{\SigmaT} S_{1}&= \frac{1}{4}b^1 \frac{\delta_2^2-\delta_1^2}{R_0}+\ord(\delta_1^3)+\ord(\delta_2^3),
\\
\frac{1}{\const}\int_{\SigmaT} S_{2}&= \frac{1}{4}b^2 \frac{\delta_2^2-\delta_1^2}{R_0}+\ord(\delta_1^3)+\ord(\delta_2^3),
\\
\frac{1}{\const}\int_{\SigmaT} S_{3}&=\frac{1}{4}b^3
\frac{\delta_2^2-\delta_1^2}{R_0}+\frac{1}{2}a\frac{\delta_2-\delta_1}{R_0}+\frac{1}{3}a^3(\delta_2^2-\delta_1^2)
+\ord(\delta_1^3)+\ord(\delta_2^3)
\end{aligned}
\label{int_Sk}
\end{equation}
%]
where
%[
\begin{align*}
\delta_1=\tau_1-\tau_0, \quad\quad \delta_2=\tau_2-\tau_0
\end{align*}
%]
and $\kappa$ is given by (\ref{def_kappa}).

Combining (\ref{int_Sk}) into a single expression and using
(\ref{def_Pdot}) and (\ref{def_Pdot_dual}) we obtain the following
expression for $\Pdot(\tau_0)\in  \textup{T}_{\c(\tau_0)}\m$
%[
\begin{align*}
\frac{1}{\const}\Pdot(\tau_0)=&\tfrac{2}{3}a^2 \tfrac{\partial}{\partial x^0} +\lim_{R_0 \rightarrow 0} \frac{1}{2R_0}a \tfrac{\partial}{\partial x^3}+ \raisebox{0.4cm}{$\displaystyle{\lim_{\substack{R_0  \rightarrow 0\\\tau_1 \rightarrow \tau_0\\\tau_2 \rightarrow \tau_0}}}$} \Big( \frac{\tau_1+\tau_2-2\tau_0}{4 R_0}\Big)b^j \tfrac{\partial}{\partial x^j} +\ord\big(\delta_1^2\big)+\ord\big(\delta_2^2\big).
\end{align*}
%]
Hence from (\ref{def_lambda}) and (\ref{def_x_i})
%[
\begin{equation}
\begin{aligned}
\frac{1}{\const}\Pdot(\tau_0)&=\tfrac{2}{3}g\big(\cddot(\tau_0),
\cddot(\tau_0)\big)\cdot(\tau_0) + \lim_{R_0 \rightarrow 0}
\frac{1}{2R_0}\cddot(\tau_0)+\lambda\cdddot(\tau_0)
+\ord\big(\delta_1^2\big)+\ord\big(\delta_2^2\big).
\end{aligned}
\label{Pdot_res_expan}
\end{equation}
%]
The first term in (\ref{Pdot_res_expan}) is the standard radiation
reaction term and the second term is the singular term to be renormalized.
The third term is proportional to $\cdddot(\tau_0)$ and therefore may
be recognised as the Schott term providing the coefficient is well
defined in the limit.

If $\lambda$ is chosen to be finite it follows immediately that all
higher order terms in the series vanish. This is because $R_0^{-1}$ is
the most divergent power of $R_0$ appearing in the
series. Mathematically we are free to choose $\lambda$ to diverge, in
which case higher order terms could be made finite. However this would require
extra renormalization in order to accommodate the $\lambda$ terms and
the resulting equation of motion would not resemble the
Lorentz-Abraham-Dirac equation.

Choosing $\lambda$ to be finite yields for
$\Pdot_{\text{EM}}(\tau)\in  \textup{T}_{\c(\tau)}\m$
%[
\begin{align}
\frac{1}{\const}\dot{\mathrm{P}}_{\text{EM}} &= \tfrac23 g(\cddot, \cddot)\cdot +\lambda \cdddot +\lim_{R_0 \rightarrow 0}\frac{1}{2 R_0}\cddot.
\end{align}
%]
The value of $\lambda$ may be fixed by satisfying the orthogonality
condition (\ref{g_Cd_Cdd}),
%[
\begin{align}
0=\frac{1}{\const}g(\Pdot_{\text{EM}}, \cdot)&=-\tfrac23g(\cddot, \cddot) +\lambda g(\cdddot, \cdot)=-(\tfrac23+\lambda )g(\cddot, \cddot).
\end{align}
%]
Therefore $\lambda=-\tfrac{2}{3}$ and the final covariant expression
for $f_{\textup{self}}$ is given by
%[
\begin{align}
f_{\textup{self}}=-\Pdot_{\textup{EM}} &= \tfrac23\kappa \big(\cdddot-g(\cddot, \cddot)\big)\cdot-\lim_{R_0 \rightarrow 0}\frac{\kappa}{2 R_0}\cddot,
\end{align}
%]
which is identical to (\ref{f_self}). Thus the complete self force has been obtained without the addition of an ad hoc term to the non-electromagnetic momentum of the particle.

%%%%%%%%%%%%%%%%%%%%%%%%%%%%%%%%%%%%%%%%%%%%%%%%%%%%%%%%%%%%%%%%%%%%%%
\section*{Conclusion}
It has been shown the complete self force may be obtained directly from the electromagnetic stress-energy-momentum tensor when using the Bhabha tube as the domain of integration. This eliminates the need to introduce the extra  ad hoc term in \eqref{P_inc_Schott}. It also proves the reason for the missing term in previous calculations is the procedure followed in taking the limits, and not the nature of the coordinates used as proposed by Gal'tsov and Spirin \cite{Galtsov02}. We have seen that a requirement for the term to appear is that the ratio of limits $\lambda$, which describes the way in which the Bhabha tube is collapsed onto the worldline, is made finite. This is a natural choice because it demands $\delta_1$, $\delta_2$ and $R_0$ to be the same order of magnitude. The specific value $\lambda=-\tfrac23$ is fixed by the orthogonality condition (\ref{g_Cd_Cdd}), however the physical justification for imposing this particular geometry on the Bhabha tube is currently unknown.

\section*{Acknowledgements}

The authors would like to thank Robin Tucker and David Burton, physics
department, Lancaster University and Adam Noble, physics department,
Strathclyde University for essential discussion.  The authors are
grateful to support from the Cockcroft Institute (STFC ST/G008248/1)
and the alpha X project, Strathclyde.

%% \nocite{Rohrlich65}
%% \nocite{vanWeert74}
%% \nocite{Rohrlich50}
%% \nocite{Rohrlich97}

\bibliography{Schott_titles}
\end{document}